*Review*

# Slow Neutron-Capture Process: Low-mass AGB stars and presolar silicon carbide grains

Nan Liu [1,2*], Sergio Cristallo[3,4], and Diego Vescovi[5]

1 Department of Physics, Washington University in St. Louis, MO 63130, USA
2 McDonnell Center for the Space Sciences, MO 63130, USA
3 INAF, Observatory of Abruzzo, 64100 Teramo, Italy
4 INFN, Section of Perugia, 06123 Perugia, Italy
5 Institute for Applied Physics, Goethe University, 60438 Frankfurt, Germany
* Correspondence: nliu@physics.wustl.edu

**Abstract:** Presolar grains are microscopic dust grains that formed in the stellar winds or explosions of ancient stars that died before the formation of the solar system. The majority (~90% in number) of presolar silicon carbide (SiC) grains, including types mainstream (MS), Y, and Z, came from low-mass C-rich asymptotic giant branch (AGB) stars, which is supported by the ubiquitous presence of SiC dust observed in the circumstellar envelope of AGB stars and the signatures of slow neutron-capture process preserved in these grains. Here, we review the status of isotope studies of presolar AGB SiC grains with an emphasis on heavy-element isotopes and highlight the importance of presolar grain studies for nuclear astrophysics. We discuss the sensitives of different types of nuclei to varying AGB stellar parameters and how their abundances in presolar AGB SiC grains can be used to provide independent, detailed constraints on stellar parameters, including $^{13}C$ formation, stellar temperature, and nuclear reaction rates.

**Keywords:** circumstellar matter 1; meteorites, meteors, meteoroids 2; nuclear reactions, nucleosynthesis, abundances 3; stars: AGB and post-AGB 4; stars: carbon 5

## 1. Introduction

In 1957, B²FH and Cameron [1, 2] proposed that elements heavier than Fe are made in stars by three nucleosynthesis processes, including slow neutron-capture process (*s*-process), rapid neutron-capture process (*r*-process), and proton-capture process (*p*-process). Since then, an important research theme in the field of nuclear astrophysics is to identify the stellar sites of the *s*-, *r*-, and *p*-processes. While the stellar sites of the *r*- and *p*-processes are still hotly debated [3-5], the detection of Tc absorption lines in Mira-type variable stars by astronomer Paul Merrill [6] pointed out that $^{99}$Tc, with a half-lifetime of 0.21 Ma, is freshly made in the stellar interior by the *s*-process operating along the valley of beta stability, thus linking the *s*-process nucleosynthesis[1] to low-mass asymptotic giant branch (AGB) stars. Sophisticated stellar models have been developed by different groups [7-9] to describe the evolution of AGB stars and associated stellar nucleosynthesis, with special attentions given to the *s*-process. According to AGB stellar models, it is recognized that the *s*-process operates in the He-intershell of low-mass (~1.5 $M_\odot$ ≤ M ≤ 3-4 $M_\odot$) AGB stars [10]. During the interpulse phase, the *s*-process is powered by a main neutron source, the $^{13}$C($\alpha$,n)$^{16}$O reaction, at a neutron density of ~$10^7$-$10^8$ cm$^{-3}$ on the timescale of 5-20 ka. As shell H-burning proceeds, the He-intershell is heated and compressed so that a thermal pulse (TP) is triggered when the temperature and density are high enough. During a TP, *s*-process products are further modified by neutron capture as the $^{22}$Ne($\alpha$,n)$^{25}$Mg reaction – a minor neutron source for the *s*-process – becomes partially activated in the He-intershell, providing neutrons at a density of $10^9$-$10^{10}$ cm$^{-3}$ on the timescale of a few years. The high-density neutron exposure produced by the minor neutron source controls the production of nuclei affected by *s*-process branch points, at which neutron-capture rates are comparable to beta-decay rates [11, 12].

Despite the abovementioned general consensus among AGB models, parameters in AGB stellar models such as mass loss rate and uncertain nuclear reaction rates such as the $^{13}$C($\alpha$,n)$^{16}$O reaction rate [13], contribute to uncertainties in model predictions for the *s*-process [12, 14]. In particular, the formation of the major neutron source $^{13}$C in the He-intershell is a key fundamental problem that is directly related to the *s*-process nucleosynthesis. The formation of $^{13}$C in the He-intershell requires the occurrence of $^{12}$C(p,$\gamma$)$^{13}$N($\beta^+$)$^{13}$C reaction chain and, in turn, a partial mixing of H from the convective envelope border into the underlying He-intershell, which is initially H free but contains abundant $^{12}$C produced by He-burning. Note that the so-called $^{13}$C pocket refers to the top thin He-intershell region that contains the major neutron source $^{13}$C. Over the last several decades, multiple physical mechanisms have been proposed to account for the partial mixing of H into the underlying He-intershell, including overshooting [8, 15-18], gravity waves [19], rotation-induced instabilities [20, 21], and magnetic buoyancy [22, 23]. It is still a hot debate in the community which is the mechanism primarily responsible for the partial mixing of H. Below, we will highlight the unique role of the heavy-element isotope data of presolar SiC grains from AGB stars in constraining AGB stellar parameters and recent progress in this research area.

## 2. Presolar Grains and In Situ Isotope Analyses

### 2.1. Presolar SiC Grains from Low-mass C-rich AGB Stars

AGB stars experience strong stellar winds and thus significant mass losses from the surface ($10^{-8}$–$10^{-5}$ $M_\odot$/year) (see [24] for a review). As hot gas is lost from the stellar surface and cools down, various dust components start to condense out of the gas when temperature drops below ~2500 K and SiC dust starts to condense at temperature below 2000 K [25]. AGB stellar models predict that stars with initial masses between ~1.5–3-4 $M_\odot$ can eventually become C-rich in the envelope during the AGB phase as during third dredge-up (TDU) events the bottom of the convective envelope penetrates into the underlying He-intershell and brings newly synthesized materials, including the He-burning product $^{12}$C and *s*-process products, to the surface. SiC dust has been observed to be present in the circumstellar envelopes of C-rich AGB stars according to a solid-state emission band at ~11.3 μm [26, 27]. AGB dust grains and their gas are lost from the surface, enter the interstellar medium (ISM), and become ISM components. Due to compression, a molecular cloud forms in a dense ISM region, and later stars like the Sun form as parts of the molecular cloud collapse. Thus, the solar system incorporated stardust grains formed in the stellar winds of ancient stars that died before the formation of the solar system.

In extraterrestrial materials from small solar system bodies such as primitive asteroids (i.e., asteroids that have never experienced any significant heating since their formation), dust grains from stars that died before the solar system formation (see Fig. 1 for examples) are preserved and identified by their exotic isotopic compositions, reflecting the compositions of their parent stars (see [28, 29] for reviews). As these stardust grains formed before the solar system formation, they are known as presolar grains. Various presolar phases from multiple stellar sources have been identified in primitive extraterrestrial materials, reflecting diverse physicochemical condensation environments among stars. Regarding SiC, ancient AGB stars are inferred to be the dominant stellar source with small contributions from core-collapse Type II supernovae [30, 31], J-type carbon stars [32], and maybe born-again AGB stars [32, 33] and novae [34-36].

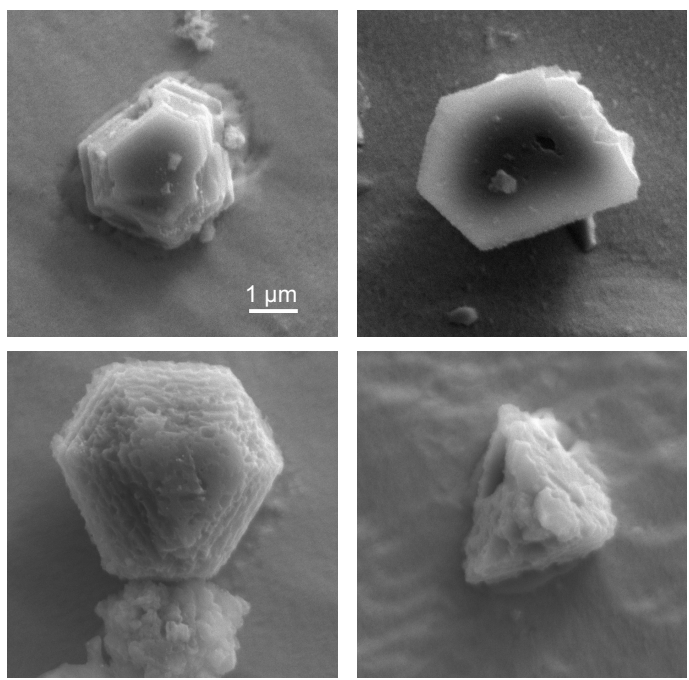

*Figure 1. Scanning electron microscopic images of presolar SiC grains extracted from CM2 chondrite Murchison using the CsF acid dissolution method described in [37].*

Presolar SiC grains from AGB stars include mainstream (MS, ~85-90%), Y (~1-3%), and Z (~1-3%) grains, which are classified based on C and Si isotope ratios (Fig. 2). MS grains are estimated to have come from AGB stars with initial masses of ~1.5–3-4 $M_\odot$, mainly because stellar models predict that such stars can become C-rich in the envelope during the AGB phase so that SiC can form in their stellar winds [7-9]. The initial metallicities of the parent stars of MS grains are more ambiguous. Given that MS grains mostly have higher-than-solar $^{29}$Si/$^{28}$Si and $^{30}$Si/$^{28}$Si ratios (Fig. 2) and that both isotope ratios are expected to increase with increasing metallicity during Galactic chemical evolution (GCE) [38], it implies that MS grains came dominantly from higher-than-solar metallicity AGB stars. Chemical heterogeneities in the Galaxy, i.e., heterogenous GCE, however, could have blurred a simple, positive correlation between the Si isotope ratios and metallicity expected from homogenous GCE [39, 40], in which case the Si isotope ratios of MS grains cannot be used to directly infer the initial metallicities of their parent stars. The current consensus in the community is that MS grains came from low-mass (~1.5–3-4 $M_\odot$), close-to-solar-metallicity (>0.5–2.0 $Z_\odot$) AGB stars [40-43]. For a long time, types Y and Z grains were thought to have come from lower-metallicity (e.g., 0.3–0.5 $Z_\odot$) AGB stars than MS grains based on light-element isotope ratios, i.e., C, Si, and Ti isotope ratios [44, 45]. Recent studies of heavy-element isotopic compositions of Y and Z grains result in a conundrum regarding the stellar origins of Y and Z grains: while the higher $^{88}Sr/^{87}Sr$ and $^{138}Ba/^{136}Ba$ ratios observed in Y and Z grains (compared to MS grains) support the lower-metallicity stellar origins of the two rare types [46], Y and Z grains exhibit Mo isotopic compositions indistinguishable from MS grains, in contrast to varying Mo isotopic patterns predicted by stellar nucleosynthesis models for AGB stars with varying stellar metallicities [47]. Also, recent statistical investigations based on natural clustering analysis techniques suggest that the classifications of MS, Y, and Z grains are quite arbitrary and not statistically significant [48, 49]. Thus, the stellar origins of Y and Z grains remain ambiguous. Finally, the indistinguishable, wide range of $^{14}N/^{15}N$ ratios observed in types MS, Y, and Z grains result mainly from contamination with terrestrial and asteroidal materials [50]. The intrinsic N isotopic compositions of AGB dust grains are characterized by large $^{14}N$ excesses compared to the solar composition [50]. It remains a question whether "uncontaminated" MS, Y, and Z grains have $^{14}N/^{15}N$ ratios [51].

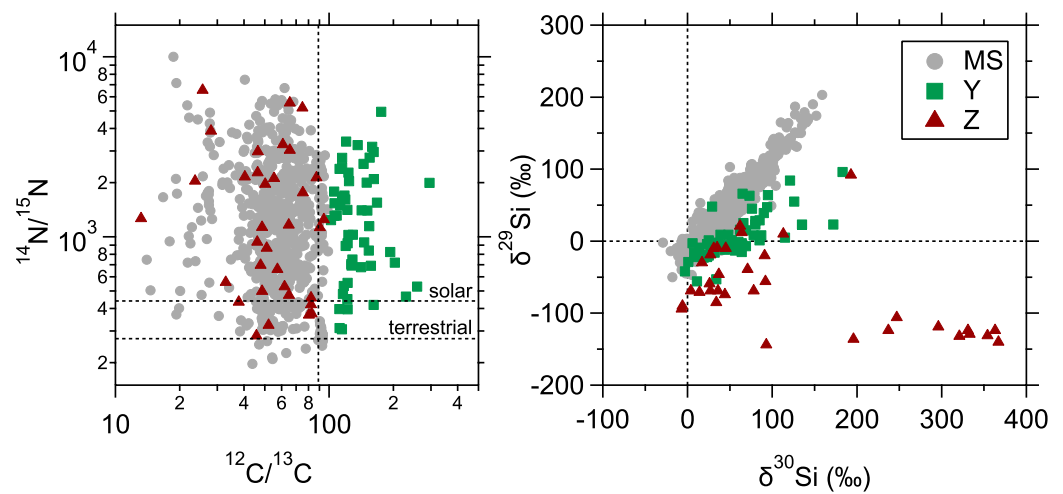

*Figure 2. Nitrogen, C, and Si isotopic ratios of presolar SiC grains from low-mass C-rich AGB stars, including types MS, Y, and Z grains [52]. The Si isotope ratios are expressed in δ notation, which is calculated using the equation $\delta^i Si = [(^i Si/^{28}Si)_{grain}/(^i Si/^{28}Si)_{solar} - 1] \times 1000‰$. Unless noted otherwise, the dashed lines represent the terrestrial composition.*

## 2.2. In situ Isotope Analysis of Presolar SiC Grains

### 2.2.1. NanoSIMS and Isotope Analyses of Light Elements

Nanoscale secondary ion mass spectrometry (NanoSIMS) is one of the primary tools for *in situ* analysis of presolar grains, especially for light-element isotope analyses [53]. Current Cameca NanoSIMS 50/50L instruments are equipped with two ion sources, $Cs^+$ and $O^-$, thus allowing the analysis of both electronegative, e.g., $C^-$, and electropositive, e.g., $Ti^+$, elements. The recent invention of a Hyperion radio-frequency ion source allows producing an $O^-$ beam of reduced size, thus enabling isotope imaging of both electronegative and electropositive elements at comparable spatial resolutions (on the order of ~100 nm at a few pA). Recent studies [50, 54-56] have shown that such high-resolution imaging is critical for excluding contamination from grain rims and adjacent grains, thus allowing for obtaining intrinsic isotopic signatures of their ancient parent stars. Depending on the instrument model, a Cameca NanoSIMS instrument is equipped with either five (NanoSIMS 50) or seven (NanoSIMS 50L) detectors, allowing simultaneous detection of five or seven ion species, respectively. In SIMS, ions of different masses are separated using a combination of an electrostatic analyzer and a magnetic analyzer. For presolar SiC, isotopes of elements like C, N, Si, S, Mg-Al, Ca-Ti, and Ti-V are routinely measured using NanoSIMS.

2.2.2. RIMS and Isotope Analyses of Heavy Elements

Isotope analyses of heavy elements face two main challenges, low concentration and isobaric interference. For instance, the solar abundance of Ba is ~200,000 times lower than that of Mg [57], and the analysis of Ba isotopes may suffer from isobaric interferences at masses 134 u ($^{134}$Ba with $^{134}$Xe), 136 u ($^{136}$Ba with $^{136}$Xe), and 138 u ($^{138}$Ba with $^{138}$La). Assuming that the Ba concentration in presolar SiC is 10 ppm [58], it translates to ~$10^5$ Ba atoms within a 1 μm grain, distributed over seven isotopes. Thus, heavy-element isotope analysis of microscopic presolar grains requires the use of a mass spectrometric technique with both high efficiency and selectivity, which points to resonance ionization mass spectrometry (RIMS) thanks to its resonant ionization capability. In contrast to SIMS, RIMS is a type of time-of-flight (ToF) mass spectrometry and targets neutral atoms instead of secondary ions released from a sample. To ionize isotopes of an element of interest in the cloud of released neutrals, RIMS shines post-ionization lasers onto the cloud with their wavelengths tuned to match the atomic structure of the element of interest. As resonant ionization is more efficient by orders of magnitude than non-resonant ionization, atoms of the element of interest are efficiently ionized while potential isobaric interferences are greatly suppressed. So far, RIMS instruments have mainly utilized ultraviolet laser beams to desorb atoms from presolar SiC, and the desorption laser beams are 1 to a few μm in size. RIMS analysis is conducted in spot mode, and imaging analysis is not yet available. For presolar SiC, isotopes of Ti, Fe, Ni, Sr, Zr, Mo, Ru, Ba, and Nd [33, 59-67] have been measured using several RIMS instruments, including CHARISMA, CHILI, and LION [59, 68, 69]. In principle, the combination of the NanoSIMS and RIMS instruments allows analyzing isotopes of most elements (except for noble gases) in the periodic table in μm-sized presolar grains.

**3. Isotope versus Element Abundances**

We first briefly describe the details of the *s*-process as laid out in the seminal papers of [1, 2]. For the *s*-process nucleosynthesis, we have

$$\frac{dN^i}{dt} = -n_n <\sigma^i v>^i N^i + n_n <\sigma^{i-1} v>^{i-1} N^{i-1} \quad (1),$$

in which, $N^i$ denotes the *s*-process abundances of a nuclide with mass *i*, $n_n$ neutron density, $\sigma^i$ the (n,γ) cross section of the nuclide *i*, and $v$ the relative neutron velocity.

Then we define two terms, neutron exposure $\tau$ [70] – an evolutionary parameter for the *s*-process – and Maxwellian-averaged neutron capture cross-section $\sigma_i^{MACS}$ as follows,

$$\tau = \int n_n v_{th} \, dt \quad (2)$$

$$\sigma_{MACS}^i = \frac{<\sigma^i v>}{v_{th}} \quad (3),$$

in which, $v_{th}$ is the thermal velocity.

By substituting $dt$ and $\sigma_{MACS}^i$ in Equation (1) with their expressions in Equation (2) and (3), respectively, we obtain

$$\frac{dN^i}{d\tau} = -\sigma_{MACS}^i N^i + \sigma_{MACS}^{i-1} N^{i-1} \quad (4).$$

Thus, if the *s*-process achieves a steady state, $\frac{dN^i}{d\tau}$ becomes zero so that we obtain

$$\sigma_{MACS}^i N^i = \sigma_{MACS}^{i-1} N^{i-1} \quad (5).$$

Equation (5) explains the observation that in regions between nuclei with magic number of neutrons, the $\sigma_{MACS}^i N^i$ product for the solar system *s*-process isotopes remains approximately constant (see Fig. 2 of [12]). The *s*-process can approximately reach a steady state in this case because it occurs on a long timescale due to slow neutron capture rates at low neutron densities ($10^7$-$10^8$ cm$^{-3}$). However, for magic nuclei, since the *s*-process cannot achieve a steady state due to their small neutron capture cross sections, $\frac{dN^i}{d\tau} \neq 0$. Thus, according to Equation (1), the *s*-process production of magic nuclei depends on neutron density and thus on the distribution of the major neutron source $^{13}$C (i.e., varying $^{13}$C density with stellar radius) in the He-intershell (see examples in Section 4.2). The study of [71] showed that the classical approach based on an exponential distribution of neutron exposures, which was first introduced by [70] and [72], overproduces the solar abundance of $^{142}$Nd (a magic nuclide), which, however, is well reproduced by AGB stellar models by adopting the same set of nuclear reaction rates. Thus, the *s*-process productions of magic nuclei have complex dependences on neutron density and need to be investigated in detail by AGB stellar nucleosynthesis models, which couple stellar models with a full nuclear reaction network to calculate *s*-process yields as a function of time (e.g., [8]) and thus differ from the classical approach that adopts a steady *s*-process flow [70, 72].

Such bottleneck effects at magic nuclei along the *s*-process path, resulting from their stable nuclear structures, explain the fact that the solar system pattern is characterized by three *s*-process peaks at $^{88}$Sr ($N_n = 50$, in which $N_n$ is the number of neutrons), $^{138}$Ba ($N_n = 82$), and $^{208}$Pb ($N_n = 126$, doubly magic with the number of protons equal to 82). The three magic nuclei $^{88}$Sr, $^{138}$Ba, and $^{208}$Pb act as the most important bottlenecks along the *s*-process path, resulting in accumulation of neutrons at these mass regions to boost their own *s*-process productions. [hs/ls] is defined as the ratio of heavy *s*-elements (e.g., Ba) at the 2nd *s*-process peak to light *s*-elements (e.g., Sr) at the 1st *s*-process peak observed in a star, which is normalized to the respective solar ratio and in logarithmic scale, and [Pb/Fe] is similar to [hs/ls] but for the ratio of Pb to Fe elemental abundance. Because of the accumulation of neutrons and non-equilibrium effects at the three bottlenecks, [hs/ls] and [Pb/Fe] are often used for comparing stellar observations with AGB model calculations to constrain the *s*-process neutron flux [73]. Physical models for partially mixing H into the He-intershell based on different mechanisms, all contain free parameters, which can be tuned to result in varying degrees of partial mixing of H and thus varying neutron fluxes. In turn, AGB models by considering most of the proposed mechanisms[2] have been shown to be capable of reproducing existing stellar observations [8, 9, 23, 43, 74, 75], which thus cannot distinguish between the various mechanisms. Since the *s*-process nucleosynthesis operates at the level of isotopes instead of elements, [hs/ls] and [Pb/Fe] are thus degenerate information for investigating the *s*-process. It is desirable to obtain the ratios of $^{88}$Sr/$^{86}$Sr and $^{138}$Ba/$^{136}$Ba, in which the denominator isotopes are both pure *s*-process isotopes and follow Equation (5), to investigate the problem of $^{13}$C formation. The ratio of $^{208}$Pb/$^{204}$Pb in presolar grains may be complicated by the radiogenic decay of $^{232}$Th to $^{208}$Pb with a half-lifetime of 14 Ga. Based on postprocessing AGB stellar models [61] showed that the *s*-process production of $^{88}$Sr/$^{86}$Sr and $^{138}$Ba/$^{136}$Ba ratios depends strongly on the detailed distribution of $^{13}$C, e.g., the $^{13}$C-pocket size, in the He-intershell and that the Sr and Ba isotope ratios of MS grains could only be explained by AGB models that adopted large $^{13}$C pockets with low $^{13}$C densities (see Fig. 9 of [61]).

Isotope ratios of certain heavy elements have been determined for a few metal-poor stars, thanks to large line broadening effects caused by the hyperfine substructures of their odd isotopes (see [76] for a review). In Fig. 3, we compare the high-resolution spectrographic measurement [77] of Ba isotopes for a metal poor star with the Ba isotopic composition of a randomly selected grain among MS grains analyzed [33, 61, 63, 78, 79], which highlights the superb precision of presolar grain measurements. Panels a-c of Fig. 3 illustrate that compared to $^{137}$Ba/$^{136}$Ba, $^{134}$Ba/$^{136}$Ba and $^{138}$Ba/$^{136}$Ba are more variable among SiC grains because the s-process productions of $^{134}$Ba and $^{138}$Ba are affected by a branch point at $^{134}$Cs ($t_{1/2}$ = 2.1 years) and the distribution of the major neutron source $^{13}$C in the He-intershell, respectively [33]. In turn, it means that we can provide constraints on different modeling inputs using these Ba isotope ratios. In detail, $^{134}$Ba/$^{136}$Ba is sensitive to the maximum stellar temperature in the He-intershell in AGB stars mainly because the stellar β−-decay rate of $^{134}$Cs has a strong temperature dependence [80]; the AGB model predictions for the $^{137}$Ba/$^{136}$Ba and $^{135}$Ba/$^{136}$Ba data in panel b of Fig. 3 are controlled by the adopted $\sigma^{135}$MACS and $\sigma^{137}$MACS values in AGB models (see Equation (5)) and $^{138}$Ba/$^{136}$Ba can be used to probe the distribution of $^{13}$C in the He-intershell [33, 41, 61]. In summary, presolar grain measurements provide a unique opportunity to determine isotope abundances in stars at a precision that far exceeds the current capability of spectrographic measurements using state-of-the-art telescopes (see Fig. 3d). In turn, presolar grain isotope data can be used to provide the most precise and detailed constraints on modeling inputs.

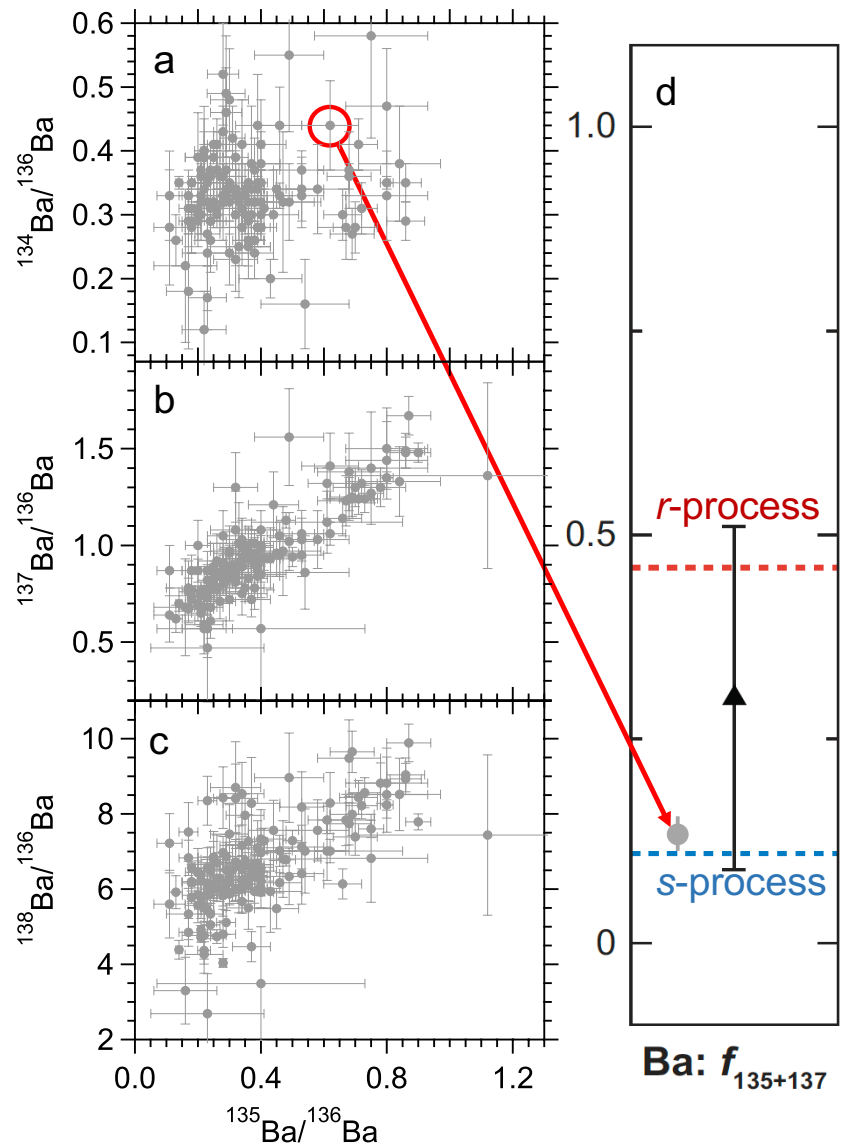

*Figure 3*. Panels a-c: RIMS Ba isotope data of MS and unclassified grains [33, 61, 63, 78, 79]; Panel d: the Ba isotope ratio of a randomly selected MS grain (highlighted by the red circle in panel a) compared to the spectrographic data of a metal-poor subgiant HD 140283 ([Fe/H]~-2.4) [77]. In panel d, $f_{135+137}$ denotes the fraction of $^{135}$Ba and $^{137}$Ba among all Ba isotopes. The dashed red and blue lines in panel d denote the pure-r- and -s-process predictions, respectively [71]. Error bars are all 1 σ.

## 4. Constraints on AGB Stellar Models from Presolar Grain Data

### 4.1. Minor Neutron Source $^{22}$Ne and Branch Points

Based on solar system *s*-process isotope abundances and AGB stellar models, [11] tested the sensitives of various *s*-process branch points to the two neutron sources and concluded that the short, high density neutron exposure produced by the $^{22}$Ne($\alpha$,n)$^{25}$Mg reaction has major effects on controlling branch points along the *s*-process path. The reaction rate of the minor neutron source $^{22}$Ne($\alpha$,n)$^{25}$Mg depends strongly on temperature and increases by 18 orders of magnitude from $1 \times 10^8$ K, typical He-intershell temperature during the interpulse phase in low-mass AGB stars, to $3 \times 10^8$ K, typical temperature in the He-intershell during TPs [81]. As the $^{22}$Ne($\alpha$,n)$^{25}$Mg reaction rate becomes increasingly higher than the rate of $^{22}$Ne($\alpha$,$\gamma$)$^{26}$Mg at T $\gtrsim 3 \times 10^8$ K, the minor neutron source $^{22}$Ne($\alpha$,n)$^{25}$Mg becomes partially activated during TPs, providing neutrons of high density to affect branch points along the *s*-process [10, 11, 82]. While *s*-process branch points are all affected by the short, high-density neutron exposure to varying degrees, certain branch points show more complex dependences on stellar temperature and/or electron density (see [11] for details). Below, we showcase one interesting branch point at $^{134}$Cs, which has a strong dependence on stellar temperature.

Cesium-134 ($t_{1/2}$ = 2 years) acts as a branch point along the *s*-process, and its $\beta^-$ decay rate increases by a factor of ~65 as temperature increases from $1 \times 10^8$ to $3 \times 10^8$ K [80]. Uncertainties in AGB model predictions for δ$^{134}$Ba are mainly controlled by uncertainties in the $\beta^-$ decay rate of $^{134}$Cs [33]. Figure 4 reveals a poor match between MS and unclassified (MS grain data hereafter) grain data and the default magnetic FRUITY (Full-network Repository of Updated Isotopic Tables & Yields) AGB models[3] (filled symbols), which are barely varied compared to the corresponding non-magnetic FRUITY models (see Fig. 6 of [33]). The similar δ$^{134}$Ba values predicted by the magnetic and nonmagnetic FRUITY AGB models, which adopted different $^{13}$C pockets (see Section 4.2 for details), demonstrate that AGB model predictions for δ$^{134}$Ba are unaffected by the adopted $^{13}$C pocket and instead controlled by the activation strength of the $^{134}$Cs branch point.

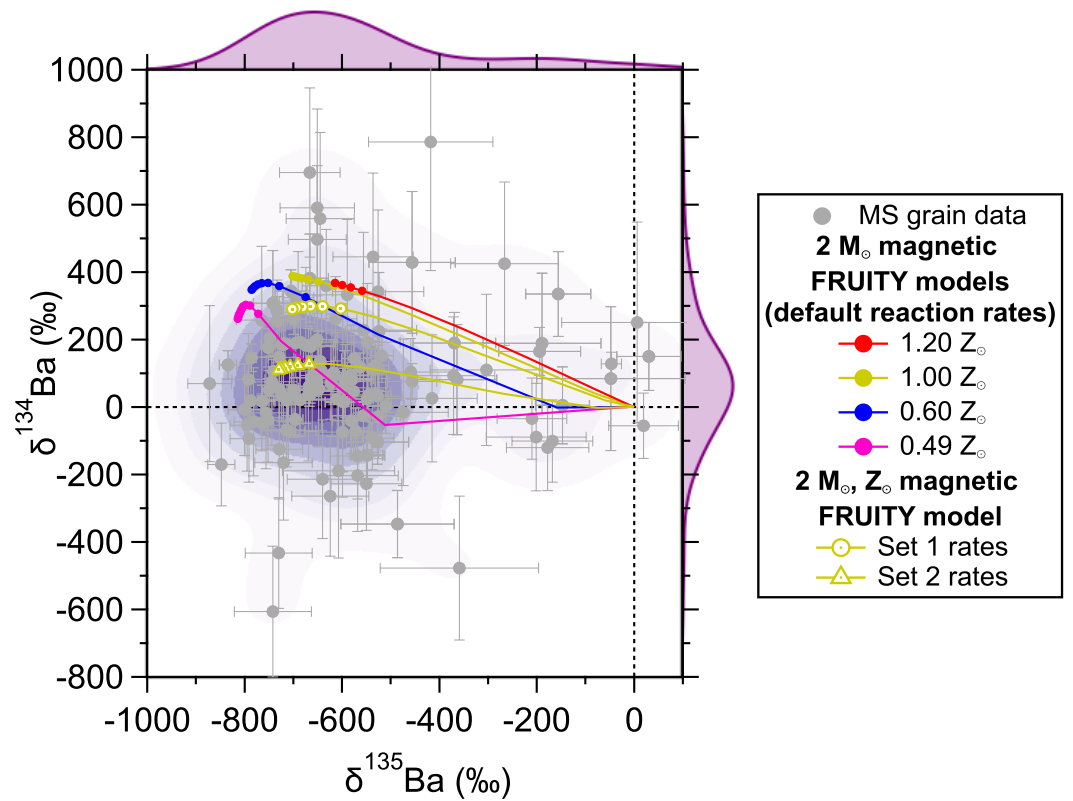

*Figure 4*. Ba 3-isotope plot comparing literature MS and unclassified grain data [61, 78] with magnetic FRUITY model calculations [75, 83]. The normalization isotope is $^{136}$Ba. A density map in linear scale is shown to illustrate the grain data distribution. For FRUITY models, lines represent O-rich phases and lines with symbols C-rich phases, during which SiC dust is expected to most likely condense [25]. The solar metallicity here refers to a solar model calibrated value of 0.0167 (see [84, 85] for details), obtained by using the present solar abundance of [52]. Error bars are all 1 σ. For all the models, we adopted the $^{22}Ne(\alpha,n)^{25}Mg$ and $^{22}Ne(\alpha,\gamma)^{26}Mg$ reaction rates recommended in [86]. Our default $^{134}Cs(\beta^-)^{134}Ba$ and $^{135}Cs(\beta^-)^{135}Ba$ reaction rates are those from [80] and default $^{134}Cs(n,\gamma)^{135}Cs$ and $^{134}Ba(n,\gamma)^{135}Ba$ reaction rates are those from [87]. For **Set 1 rates**, we adopted the $^{134}Cs(\beta^-)^{134}Ba$ and $^{135}Cs(\beta^-)^{135}Ba$ reaction rates from [88]. For **Set 2 rates**, we additionally adopted the upper limits of the $^{134}Cs(n,\gamma)^{135}Cs$ and $^{134}Ba(n,\gamma)^{135}Ba$ reaction rates from [89] and [90], respectively.

As discussed in [33], model predictions for δ$^{134}$Ba can be lowered if (*i*) the efficiency of the minor neutron source $^{22}Ne(\alpha,n)^{25}Mg$ is increased, e.g., enhanced $^{22}Ne(\alpha,n)^{25}Mg$ reaction rate and/or (*ii*) $^{134}$Cs β$^-$ decay rate is reduced. Recent experimental and theoretical studies are in favor of possibility (*ii*), because significantly lowered $^{22}Ne(\alpha,n)^{25}Mg$ rate, which was adopted in the magnetic FRUITY models in Fig. 4, is inferred at relevant low-mass AGB temperatures [86, 91] and $^{134}$Cs β$^-$ decay rate is estimated to be significantly lower [88, 92, 93] than that recommended by [80]. The study of [92] showed that adopting their newly calculated $^{134}$Cs β$^-$ decay rate reduces Monash AGB model predictions for δ$^{134}$Ba by ~300‰. The reduced $^{134}$Cs β$^-$ decay rate of [92] also allows Monash AGB model predictions for δ$^{134}$Ba to reach as low as –200‰, thus explaining several MS grains with negative δ$^{134}$Ba values. These negative δ$^{134}$Ba values were shown to be a problem for AGB models with the old $^{134}$Cs β$^-$ decay rate, based on which the grains were speculated to have come from born-again AGB stars that experienced an intermediate neutron capture process [33]. The new $^{134}$Cs β$^-$ decay rate thus provides a better explanation to the δ$^{134}$Ba values of MS grains. However, when we adopted the new $^{134}$Cs and $^{135}$Cs β$^-$ decay rates from [88] (**Set 1 rates** in Fig. 4), our 2 $M_\odot$, $Z_\odot$ magnetic FRUITY model prediction for δ$^{134}$Ba dropped by only ~100‰, likely because the maximum stellar temperatures during TPs are lower in FRUITY stellar models than in Monash stellar models. It is possible to reach the center of the grain distribution if the model also adopts the upper limits for both the $^{134}$Cs$(n,\gamma)^{135}$Cs and $^{134}$Ba$(n,\gamma)^{135}$Ba reaction rates from [89] and [90], respectively (**Set 2 rates** in Fig. 4). Thus, it remains a question whether MS grains with large negative δ$^{134}$Ba values came from low-mass AGB stars or born-again AGB stars.

Other important branch points that affect the isotope abundances of heavy elements that have been measured in MS SiC grains, include $^{85}$Kr (isomeric state, $t_{1/2}$ = 4.5 hours; ground state, $t_{1/2}$ = 11 years), $^{94}$Nb ($t_{1/2}$ = 2.0 × 10$^4$ years), $^{95}$Zr ($t_{1/2}$ = 64 days), $^{135}$Cs ($t_{1/2}$ = 2.3 × 10$^6$ years), and $^{137}$Cs ($t_{1/2}$ = 30 years). Future RIMS measurements of rare-earth-element isotope ratios in MS grains [62] will likely provide better constraints on the maximum stellar temperature in the He-intershell during TPs as there are many strong branch points along the s-process in this mass region, e.g., $^{147}$Pm ($t_{1/2}$ = 2.6 years), $^{151}$Sm ($t_{1/2}$ = 90 years), $^{152}$Eu ($t_{1/2}$ = 13.5 years).

*4.2. Formation of Major Neutron Source $^{13}$C*

Because of bottleneck effects, MS grain data for $^{88}$Sr/$^{86}$Sr and $^{138}$Ba/$^{136}$Ba can be used to probe the distribution of $^{13}$C formed in the He-intershell (see discussion in Section 3). In Fig 5., we compare MS grain data with two sets of FRUITY models for 2 $M_\odot$ AGB stars with 0.36–1.20 $Z_\odot$ that consider different mechanisms for the $^{13}$C formation.

Convective overshooting can lead to a partial mixing of H into the He-intershell as convective eddies in the envelope cross the bottom of the envelope and move downward into the He-intershell. For the non-magnetic FRUITY models in Fig. 5a, convective overshooting is considered to be primarily responsible for driving the partial mixing of H into the He intershell [8, 17], and the mixing velocity is estimated as $v(r) = v_{cb} e^{-\frac{\delta r}{\beta H_P}}$, where $v(r)$ is the velocity of envelope material injected into the He-intershell at distance $r$ from the border, $v_{cb}$ the velocity at the convective border, $\delta r$ the distance from the border, $H_P$ the pressure scale height at the border, and $\beta$ a free parameter. The β value is set to 0.1 for the non-magnetic models in Fig. 4a to maximize the efficiency of $^{13}$C formation [8].

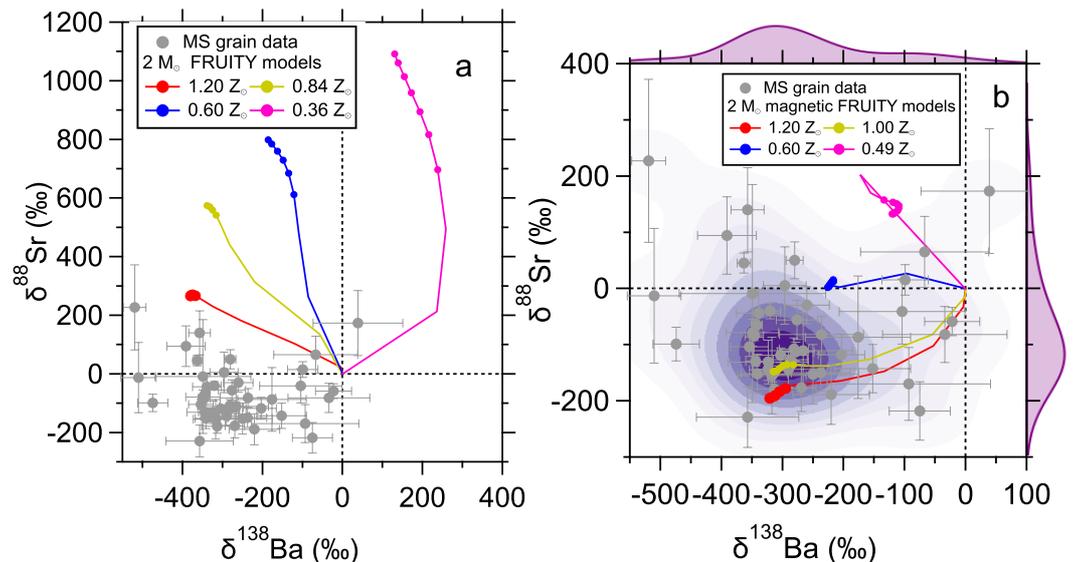

*Figure 5.* Sr and Ba isotope plots comparing MS grain data from [61, 78] with FRUITY model calculations that consider **no magnetic effects** [8, 74] in panel a and **magnetic effects** in panel b (the same as the default magnetic FRUITY models in Fig. 4). The normalization isotopes for δ$^{88}$Sr and δ$^{138}$Ba are $^{86}$Sr and $^{136}$Ba, respectively. Error bars are all 1σ. Note that the metallicities of the two sets of models are slightly different.

For the magnetic FRUITY models in Fig. 5b, magnetic buoyancy is the mechanism that drives partial mixing of H downward because of mass conservation, since uprising magnetic flux tubes carry mass flows to move upward [22, 23, 75, 93, 94]. The mixing velocity in this case is estimated as $v(r) = u_p \left(\frac{r_p}{r}\right)^{k+2}$, where $u_P$ is proportionally related to the initial velocity of magnetic flux tubes and thus the magnetic field strength $B_\varphi$, $r_P$ is the distance from the stellar center, and $k$ describes the dependence of density $\rho$ on the stellar radius, $\rho \propto r_P^k$.

It was shown in [75] that the two mechanisms abovementioned lead to different $^{13}$C distributions in the He-intershell: while in the overshooting scenario the amount of $^{13}$C drops quickly to zero in the He-intershell because the overshooting velocity follows an exponential decaying profile, magnetic buoyancy can lead to formation of $^{13}$C in deeper He-intershell with a more flattened $^{13}$C distribution given the power law dependence of the mixing velocity (see Fig. 3 of [75] for comparison). [75] further showed that while varying β values could not reach a good match between MS grain data and low-mass FRUITY AGB models, it provides a satisfying match to the MS grain data by adopting $B_\varphi$ = 5 ×10$^4$ G and $u_P$ = 5 ×10$^{-5}$ cm/s in 2 $M_\odot$ FRUITY AGB models (Fig. 4b).

In addition to the $^{13}$C distribution in the He-intershell, AGB model predictions for $\delta^{88}$Sr are also affected by a branch point at $^{85}$Kr along the *s*-process. However, we noticed negligible changes in the model predictions for $\delta^{88}$Sr when we increased the $^{22}$Ne($\alpha$,n)$^{25}$Mg reaction rate by a factor of 2–3 in the magnetic FRUITY models, thus pointing to a negligible effect of the $^{85}$Kr branch point (and thus the minor neutron source) on FRUITY AGB model predictions for $\delta^{88}$Sr. The differences in the predicted $\delta^{88}$Sr and $\delta^{138}$Ba values between the magnetic and nonmagnetic FRUITY models (which adopted slightly different $^{22}$Ne($\alpha$,n)$^{25}$Mg and $^{22}$Ne($\alpha$,$\gamma$)$^{26}$Mg reaction rates) in Fig. 5, therefore, result primarily from the different $^{13}$C pockets formed in the two sets of models.

Although it is tempting to conclude based on the data-model comparisons in Fig. 5 that magnetic buoyancy is the mechanism primarily responsible for the $^{13}$C formation, more efforts are needed in different subfields to corroborate the results of Fig. 5.

**(1) Presolar Grains**: Correlated Sr and Ba isotope analyses of more MS grains using the new generation of RIMS instruments [59, 68] are needed to better quantify the MS grain distribution. The study of [41] showed that compared to the exponential distribution by overshooting, the deeper, more flattened $^{13}$C distribution resulting from magnetic buoyancy, reduces the sensitives of $\delta^{88}$Sr and $\delta^{138}$Ba to the H mixing depth. Thus, we need more MS grain data for $\delta^{88}$Sr and $\delta^{138}$Ba to determine the data variability, which will help to assess the primary mechanism responsible for the $^{13}$C formation. A better understanding of the MS grain data distribution will also allow for a quantitative assessment of the quality of data-model comparisons, which, so far, have been conducted mainly in a qualitative way.

**(2) Nuclear Experiments**: AGB model predictions for $\delta^{88}$Sr and $\delta^{138}$Ba rely directly on the $\sigma_{MACS}$ values of $^{86}$Sr, $^{88}$Sr, $^{136}$Ba, and $^{138}$Ba. Given the small $^{88}$Sr and $^{138}$Ba $\sigma_{MACS}$ values, current AGB model uncertainties in $\delta^{88}$Sr and $\delta^{138}$Ba are controlled by uncertainties in the $^{86}$Sr (±10%) and $^{136}$Ba (±3%) $\sigma_{MACS}$ values [95][4], respectively, which correspond to ~200‰ and ~50‰ uncertainties in low-mass AGB model predictions for $\delta^{88}$Sr and $\delta^{138}$Ba, respectively [33, 61]. As the full range of $\delta^{88}$Sr values observed among MS grains is only ~400‰ (Fig. 5), new measurements of $^{86}$Sr $\sigma_{MACS}$ values are urgently needed to reduce the model uncertainty for $\delta^{88}$Sr.

**(3) Stellar Modeling**: Implementation of magnetic-buoyancy effects in other stellar codes such as NuGrid [9] is needed to test whether the effect of magnetic buoyancy on the *s*-process production is stellar code dependent. It was shown in [42, 43] that 2 $Z_\odot$ Monash models that adopt an exponentially decayed mixing profile also provide a good match to the heavy-element isotope data of MS grains. It remains to see whether adoption of the formula for magnetic buoyancy can further improve the data-model agreements for 2 $Z_\odot$ and other lower-metallicity Monash models, given uncertainties in the initial metallicities of grains' parent stars. Finally, we note that the good grain-model agreement in Fig. 5b mainly points out that the MS grain data are in favor of a deep, flattened $^{13}$C distribution in the He-intershell. More modeling efforts are needed to investigate whether magnetic buoyancy is the sole mechanism that could lead to such a distribution.

*4.3. Constraints on Neutron Capture Cross Sections*

Besides isotopes that are affected by branching effects and magic nuclei, the relative *s*-process productions of other nuclei in the heavy mass region are dominantly controlled by the adopted neutron capture cross sections according to Equation (5). Thus, the abundances of such nuclei can be measured in AGB dust grains for comparison with AGB models to examine the adopted $\sigma_{MACS}$ values. Such an example is shown in Fig. 6 for illustration. Among Mo isotopes, $^{92}$Mo and $^{94}$Mo are *p*-process isotopes with a small *s*-process contribution to $^{94}$Mo [75]. The *p*-process is an umbrella term for multiple nucleosynthesis processes that can produce proton-rich nuclei in the heavy mass region. So far, it has been proposed that the *p*-process includes gamma process in Type II core-collapse supernovae and Type Ia supernovae and neutrino-*p* process in Type II core-collapse supernovae (see [96] and references therein).

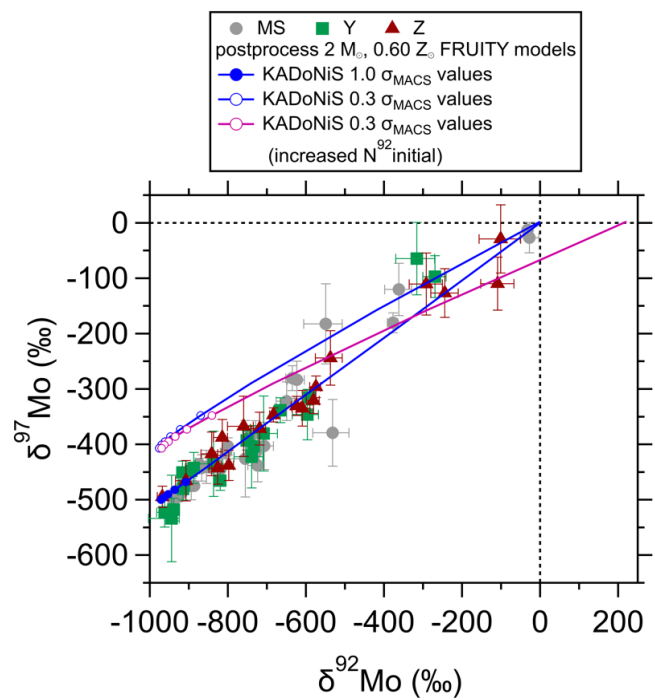

*Figure 6*. Mo 3-isotope plot comparing MS, Y, and Z grain data [47, 66] with the same FRUITY model but calculated with different Mo $\sigma_{MACS}$ values. The normalization isotope is $^{96}$Mo. Note that the FRUITY models were run in a postprocess way (see [47] for details), different from the fully coupled FRUITY models shown in Figs. 4 & 5. Error bars are all 1$\sigma$.

It was shown that AGB model predictions for Mo isotopes are unaffected by the two neutron sources and controlled by the adopted Mo $\sigma_{MACS}$ values [97]. According to Equation (5), the ratio of $^{97}$Mo/$^{96}$Mo (i.e., $N^{97}/N^{96}$) is approximately related to $\sigma^{97}_{MACS}/\sigma^{96}_{MACS}$. The *p*-process isotope $^{92}$Mo is off the *s*-process path and is, therefore, destroyed by neutron capture during the *s*-process in AGB stars. As a result, the final $^{92}$Mo abundance in the AGB envelope is dominantly controlled by the initially adopted $^{92}$Mo abundance, which is poorly known due to uncertainties in modeling the GCE of *p*-process nuclei [96]. AGB models in Fig. 6 predict extremely low $^{92}$Mo/$^{96}$Mo ratios for the final envelope composition because the pure *s*-process isotope $^{96}$Mo is significantly overproduced by a factor of ~40 in the envelope compared to its initial abundance. A comparison of the two KADoNiS 0.3 AGB models with different initial $^{92}$Mo abundances in Fig. 6 reveals that 200‰ increase in the initial $^{92}$Mo abundance leads to ~80‰ increase in $^{92}$Mo/$^{96}$Mo for the final envelope composition but only 6‰ increase in δ$^{92}$Mo due to the non-linearity of the delta notation when approaching −1000‰ (i.e., a negligible difference in the final envelope composition between the two KADoNiS 0.3 models). Thus, variation in the initial stellar composition cannot resolve the data-model discrepancy by adopting KADoNiS 0.3 σ$_{MACS}$ values. Instead, the grain data are well reproduced by adopting the Mo isotope σ$_{MACS}$ recommended by KADoNiS$^4$ 1.0 [95]. Better agreements with KADoNiS 1.0 Mo σ$_{MACS}$ values were also observed for $^{95}$Mo/$^{96}$Mo and $^{98}$Mo/$^{96}$Mo ratios. The AGB grain data thus provide strong support to the new Mo σ$_{MACS}$ values recommended by KADoNiS 1.0 (see [47] for more details). Other examples include $^{137}$Ba/$^{136}$Ba versus $^{135}$Ba/$^{136}$Ba (Fig. 3b) and $^{99,\,101,\,102}$Ru/$^{100}$Ru versus $^{104}$Ru/$^{100}$Ru, which can be used to constrain the respective σ$_{MACS}$ values [67].

Finally, we note that except for *p*- and *r*-process isotopes, other heavy-element isotopes that receive major *s*-process contributions can be significantly overproduced by the *s*-process nucleosynthesis in AGB stars; as a result, their final abundances in the envelope are dominantly controlled by the *s*-process nucleosynthesis. Taking $^{95}$Mo, $^{96}$Mo, $^{97}$Mo, and $^{98}$Mo for example, the AGB models in Fig. 6 predict that their final abundances in the envelope are increased by factors of ~20, ~40, ~25, and ~35, respectively, which means that the contributions of the initial abundances are small (<5%). Besides, the heavy-element isotopic compositions of AGB SiC grain data all seem to suggest that their parent AGB stars had close-to-solar initial heavy-element isotopic compositions (see [66] for details), although the data could alternatively imply significant asteroidal/terrestrial contamination.

## 5. Conclusions

Presolar grains are bona fide stellar materials. The discovery of presolar grains in primitive extraterrestrial materials allows for isotope analysis in the laboratory using modern mass spectrometric techniques at a precision that far exceeds what can be achieved by spectrographic measurements using state-of-the-art telescopes. In this short review, we summarized the sensitives of different types of heavy-element isotopes and how their abundances in MS grains can be used to constrain AGB stellar parameters.

Branch points exist along the *s*-process path, where neutron capture competes with beta decay. For isotopes that are affected by branch points, their isotope ratios in MS grains can be used to constrain the He-intershell temperature during TPs because (*i*) the efficiency of the $^{22}$Ne(α,n)$^{25}$Mg reaction depends strongly on the maximum temperature in the He-intershell during TPs and (*ii*) the β$^-$ decay rates of certain branch points are sensitive to temperature. One good example is the ratio of $^{134}$Ba/$^{136}$Ba, but a stringent constraint on the maximum temperature is hampered by the uncertain β$^-$ decay rate of $^{134}$Cs. The reduced $^{134}$Cs β$^-$ decay rates reported by two recent independent theoretical studies, enable a better match of AGB model calculations with MS grain data. AGB model predictions for $^{134}$Ba/$^{136}$Ba, however, still suffer from uncertainties in stellar parameters such as the maximum temperature during TPs. Thus, it remains a question whether a few MS grains with quite low $^{134}$Ba/$^{136}$Ba ratios came from low-mass AGB stars or born-again AGB stars.

Isotopes with magic numbers of neutrons have stable nuclear structures and thus extremely small neutron capture cross sections, resulting in accumulation of neutrons at these bottlenecks along the *s*-process path. We showed that the combination of $^{88}$Sr/$^{86}$Sr and $^{138}$Ba/$^{136}$Ba ratios of MS grains can be used to probe the $^{13}$C distribution in the He-intershell. Although several studies have shown that magnetic-buoyancy-induced $^{13}$C formation leads to better data-model agreements, it requires more grain data, more precise nuclear inputs, and more modeling efforts to corroborate this observation.

For *s*- and *s,r*-isotopes that are unaffected by branching and bottleneck effects, their relative *s*-process productions are barely affected by the two neutron sources and dominantly controlled by the respective $\sigma_{MACS}$ values adopted in AGB models. Thus, measuring their isotope abundances in MS grains allows examination of currently recommended $\sigma_{MACS}$ values. We showcased the Mo isotopic compositions of MS, Y, and Z grains, which are in favor of the $\sigma_{MACS}$ values recommended by KADoNiS v1.0 as compared to KADoNiS v0.3. Thus, analyses of these isotope abundances in MS grains provide a novel way to scrutinize currently recommended neutron capture cross sections at relevant AGB temperatures.

Given the multielement isotope analysis capabilities of both NanoSIMS and the new generation of RIMS instruments [59, 68], it is promising to obtain heavy-element isotope data for a larger number of MS grains (and also more Y and Z grains) and for more elements, e.g., rare-earth elements [62], and also with better precisions in the coming years. Accompanied by the synergic efforts in the communities of nuclear physics and stellar modelling, we will likely gain more insights into the evolution and stellar nucleosynthesis of low-mass C-rich AGB stars in the near future.

**Author Contributions:** Conceptualization, N. L.; methodology, N.L.; modeling, S.C. and D.V.; writing—original draft preparation, N. L.; writing—review and editing, N.L., S.C., and D.V.; visualization, N.L.; All authors have read and agreed to the published version of the manuscript.

**Funding:** This research was funded by NASA through grant 80NSSC20K0387 to N.L. D.V. acknowledges the financial support of the German-Israeli Foundation (GIF No. I-1500-303.7/2019).

**Data Availability Statement:** The presolar grain data included in this study can be found in the Presolar Grain Database available at https://presolar.physics.wustl.edu/presolar-grain-database/.

**Acknowledgments:** This review paper is dedicated to honoring the career of Maurizio Busso, a pioneer of the *s*-process stellar modeling and the inventor of magnetic-buoyancy-induced $^{13}$C formation. We thank two anonymous reviewers for their critical and helpful reviews.

**Conflicts of Interest:** The authors declare no conflict of interest.

## Notes

1. The *s*-process nucleosynthesis throughout the manuscript refers to the main *s*-process specifically. We remind the reader that there is also the weak *s*-process operating in massive stars, which can efficiently produce heavy elements before the first *s*-process peak at Sr (see [98] for details).
2. It is generally recognized that rotation-induced instabilities can modify the distribution of formed $^{13}$C but cannot act as the primary cause of partial mixing of H into the He-intershell [20, 21].
3. Non-magnetic FRUITY model calculations are available at http://fruity.oa-teramo.inaf.it/, and magnetic FRUITY model calculations are not yet available online.
4. KADoNiS stands for Karlsruhe Astrophysical database of Nucleosynthesis in Stars. The KADoNiS v0.3 and v1.0 databases are available at https://www.kadonis.org/ and https://exp-astro.de/kadonis1.0/, respectively.

## References

1. Burbidge, E.M.; Burbidge, G.R.; Fowler, W.A.; Hoyle, F. Synthesis of the elements in stars. *Rev. Mod. Phys.* **1957**, *29*, 547-650. http://doi.org/10.1103/RevModPhys.29.547.
2. Cameron, A.G.W. Nuclear reactions in stars and nucleogenesis. *Publ. Astron. Soc. Pac.* **1957**, *69*, 201. http://doi.org/10.1086/127051.
3. Tanvir, N.R.; Levan, A.J.; González-Fernández, C.; Korobkin, O.; Mandel, I.; Rosswog, S.; Hjorth, J.; D'avanzo, P.; Fruchter, A.S.; Fryer, C.L.; Kangas, T.; Milvang-Jensen, B.; Rosetti, S.; Steeghs, D.; Wollaeger, R.T.; Cano, Z.; Copperwheat, C.M.; Covino, S.; D'elia, V.; De Ugarte Postigo, A.; Evans, P.A.; Even, W.P.; Fairhurst, S.; Figuera Jaimes, R.; Fontes, C.J.; Fujii, Y.I.; Fynbo, J.P.U.; Gompertz, B.P.; Greiner, J.; Hodosan, G.; Irwin, M.J.; Jakobsson, P.; Jørgensen, U.G.; Kann, D.A.; Lyman, J.D.; Malesani, D.; Mcmahon, R.G.; Melandri, A.; O'brien, P.T.; Osborne, J.P.; Palazzi, E.; Perley, D.A.; Pian, E.; Piranomonte, S.; Rabus, M.; Rol, E.; Rowlinson, A.; Schulze, S.; Sutton, P.; Thöne, C.C.; Ulaczyk, K.; Watson, D.; Wiersema, K.; Wijers, R.a.M.J.

The emergence of a lanthanide-rich kilonova following the merger of two neutron stars. *Astrophys. J.* **2017**, *848*,L27. http://doi.org/10.3847/2041-8213/aa90b6.

4. Kajino, T.; Aoki, W.; Balantekin, A.B.; Diehl, R.; Famiano, M.A.; Mathews, G.J. Current status of r-process nucleosynthesis. *Prog. Part. Nucl. Phys.* **2019**, *107*,109-166. http://doi.org/10.1016/j.ppnp.2019.02.008.

5. Arnould, M.; Goriely, S. The p-process of stellar nucleosynthesis: astrophysics and nuclear physics status. *Phys. Rep.* **2003**, *384*,1-84. http://doi.org/10.1016/s0370-1573(03)00242-4.

6. Merrill, P.W. Emission lines in the spectra of long-period variable stars. *J. R. Astron. Soc. Can.* **1952**, *46*,181.

7. Karakas, A.I.; Lugaro, M. Stellar yields from metal-rich asymptotic giant branch models. *Astrophys. J.* **2016**, *825*,26. http://doi.org/10.3847/0004-637x/825/1/26.

8. Cristallo, S.; Straniero, O.; Gallino, R.; Piersanti, L.; Domínguez, I.; Lederer, M.T. Evolution, nucleosynthesis, and yields of low-mass asymptotic giant branch stars at different metallicities. *Astrophys. J.* **2009**, *696*,797-820. http://doi.org/10.1088/0004-637x/696/1/797.

9. Battino, U.; Tattersall, A.; Lederer-Woods, C.; Herwig, F.; Denissenkov, P.; Hirschi, R.; Trappitsch, R.; Den Hartogh, J.W.; Pignatari, M.; Collaboration, N. NuGrid stellar data set - III. updated low-mass AGB models and s-process nucleosynthesis with metallicities Z= 0.01, Z = 0.02, and Z = 0.03. *Mon. Not. R. Astron. Soc.* **2019**, *489*,1082-1098. http://doi.org/10.1093/mnras/stz2158.

10. Gallino, R.; Busso, M.; Picchio, G.; Raiteri, C.M.; Renzini, A. On the role of low-Mass asymptotic giant branch stars in producing a solar system distribution of *s*-process isotopes. *Astrophys. J.* **1988**, *334*,L45. http://doi.org/10.1086/185309.

11. Bisterzo, S.; Gallino, R.; Käppeler, F.; Wiescher, M.; Imbriani, G.; Straniero, O.; Cristallo, S.; Görres, J.; Deboer, R.J. The branchings of the main s-process: their sensitivity to $\alpha$-induced reactions on $^{13}$C and $^{22}$Ne and to the uncertainties of the nuclear network. *Mon. Not. R. Astron. Soc.* **2015**, *449*,506-527. http://doi.org/10.1093/mnras/stv271.

12. Käppeler, F.; Gallino, R.; Bisterzo, S.; Aoki, W. The *s* process: Nuclear physics, stellar models, and observations. *Rev. Mod. Phys.* **2011**, *83*,157-194. http://doi.org/10.1103/RevModPhys.83.157.

13. Cristallo, S.; La Cognata, M.; Massimi, C.; Best, A.; Palmerini, S.; Straniero, O.; Trippella, O.; Busso, M.; Ciani, G.F.; Mingrone, F.; Piersanti, L.; Vescovi, D. The Importance of the $^{13}$C($\alpha$,n)$^{16}$O reaction in asymptotic giant branch stars. *Astrophys. J.* **2018**, *859*,105. http://doi.org/10.3847/1538-4357/aac177.

14. Herwig, F. Evolution of asymptotic giant branch stars. *Annu. Rev. Astron. Astrophys.* **2005**, *43*,435-479. http://doi.org/10.1146/annurev.astro.43.072103.150600.

15. Freytag, B.; Ludwig, H.-G.; Steffen, M. Hydrodynamical models of stellar convection. The role of overshoot in DA white dwarfs, A-type stars, and the Sun. *Astron. Astrophys.* **1996**, *313*,497-516.

16. Herwig, F.; Bloecker, T.; Schoenberner, D.; El Eid, M. Stellar evolution of low and intermediate-mass stars. IV. Hydrodynamically-based overshoot and nucleosynthesis in AGB stars. *Astron. Astrophys.* **1997**, *324*,L81-L84.

17. Straniero, O.; Gallino, R.; Cristallo, S. s process in low-mass asymptotic giant branch stars. *Nucl. Phys. A* **2006**, *777*,311-339. http://doi.org/10.1016/j.nuclphysa.2005.01.011.

18. Battino, U.; Pignatari, M.; Ritter, C.; Herwig, F.; Denisenkov, P.; Den Hartogh, J.W.; Trappitsch, R.; Hirschi, R.; Freytag, B.; Thielemann, F.; Paxton, B. Application of a theory and simulation-based convective boundary mixing model for AGB star evolution and nucleosynthesis. *Astrophys. J.* **2016**, *827*,30. http://doi.org/10.3847/0004-637x/827/1/30.

19. Denissenkov, P.A.; Tout, C.A. Partial mixing and formation of the $^{13}$C pocket by internal gravity waves in asymptotic giant branch stars. *Mon. Not. R. Astron. Soc.* **2003**, *340*,722-732. http://doi.org/10.1046/j.1365-8711.2003.06284.x.

20. Herwig, F.; Langer, N.; Lugaro, M. The s-process in rotating asymptotic giant branch stars. *Astrophys. J.* **2003**, *593*,1056-1073. http://doi.org/10.1086/376726.

21. Piersanti, L.; Cristallo, S.; Straniero, O. The effects of rotation on s-process nucleosynthesis in asymptotic giant branch stars. *Astrophys. J.* **2013**, *774*,98.   http://doi.org/10.1088/0004-637x/774/2/98.

22. Nucci, M.C.; Busso, M. Magnetohydrodynamics and deep mixing in evolved stars. I. two- and three-dimensional analytical models for the asymptotic giant branch. *Astrophys. J.* **2014**, *787*,141.   http://doi.org/10.1088/0004-637x/787/2/141.

23. Trippella, O.; Busso, M.; Palmerini, S.; Maiorca, E.; Nucci, M.C. s-processing in AGB stars revisited. II. Enhanced $^{13}$C production through MHD-induced mixing. *Astrophys. J.* **2016**, *818*,125.   http://doi.org/10.3847/0004-637x/818/2/125.

24. Höfner, S.; Olofsson, H. Mass loss of stars on the asymptotic giant branch. Mechanisms, models and measurements. *Annu. Rev. Astron. Astrophys.* **2018**, *26*,1.   http://doi.org/10.1007/s00159-017-0106-5.

25. Lodders, K.; Fegley, B., Jr. The origin of circumstellar silicon carbide grains found in meteorites. *Meteoritics* **1995**, *30*,661. http://doi.org/10.1111/j.1945-5100.1995.tb01164.x.

26. Speck, A.K.; Corman, A.B.; Wakeman, K.; Wheeler, C.H.; Thompson, G. Silicon carbide absorption features: Dust formation in the outflows of extreme carbon stars. *Astrophys. J.* **2009**, *691*,1202-1221.   http://doi.org/10.1088/0004-637x/691/2/1202.

27. Speck, A.K.; Thompson, G.D.; Hofmeister, A.M. The effect of stellar evolution on SiC dust grain sizes. *Astrophys. J.* **2005**, *634*,426-435.   http://doi.org/10.1086/496955.

28. Zinner, E. 2014. "Presolar grains." In *Meteorites and Cosmochemical Processes*, edited by Davis, Andrew M., 181-213. Oxford: Elsevier.

29. Nittler, L.R.; Ciesla, F. Astrophysics with extraterrestrial materials. *Annu. Rev. Astron. Astrophys.* **2016**, *54*,53-93. http://doi.org/10.1146/annurev-astro-082214-122505.

30. Nittler, L.R.; Amari, S.; Zinner, E.; Woosley, S.E.; Lewis, R.S. Extinct $^{44}$Ti in presolar graphite and SiC: proof of a supernova origin. *Astrophys. J.* **1996**, *462*,L31.   http://doi.org/10.1086/310021.

31. Liu, N.; Nittler, L.R.; O'd. Alexander, C.M.; Wang, J. Late formation of silicon carbide in type II supernovae. *Sci. Adv.* **2018**, *4*,eaao1054.   http://doi.org/10.1126/sciadv.aao1054.

32. Liu, N.; Stephan, T.; Boehnke, P.; Nittler, L.R.; O'd. Alexander, C.M.; Wang, J.; Davis, A.M.; Trappitsch, R.; Pellin, M.J. J-type carbon stars: a dominant source of $^{14}$N-rich presolar SiC grains of type AB. *Astrophys. J.* **2017**, *844*,L12. http://doi.org/10.3847/2041-8213/aa7d4c.

33. Liu, N.; Savina, M.R.; Davis, A.M.; Gallino, R.; Straniero, O.; Gyngard, F.; Pellin, M.J.; Willingham, D.G.; Dauphas, N.; Pignatari, M.; Bisterzo, S.; Cristallo, S.; Herwig, F. Barium isotopic composition of mainstream silicon carbides from Murchison: constraints for *s*-process nucleosynthesis in asymptotic giant branch stars. *Astrophys. J.* **2014**, *786*,66. http://doi.org/10.1088/0004-637x/786/1/66.

34. Amari, S.; Gao, X.; Nittler, L.R.; Zinner, E.; José, J.; Hernanz, M.; Lewis, R.S. Presolar grains from novae. *Astrophys. J.* **2001**, *551*,1065-1072.   http://doi.org/10.1086/320235.

35. Nittler, L.R.; Hoppe, P. Are presolar silicon carbide grains from novae actually from supernovae? *Astrophys. J.* **2005**, *631*,L89-L92.   http://doi.org/10.1086/497029.

36. Liu, N.; Nittler, L.R.; O'd. Alexander, C.M.; Wang, J.; Pignatari, M.; José, J.; Nguyen, A. Stellar origins of extremely $^{13}$C- and $^{15}$N-enriched presolar SiC grains: novae or supernovae? *Astrophys. J.* **2016**, *820*,140.   http://doi.org/10.3847/0004-637x/820/2/140.

37. Nittler, L.R.; O'd. Alexander, C.M. Automated isotopic measurements of micron-sized dust: application to meteoritic presolar silicon carbide. *Geochim. Cosmochim. Acta* **2003**, *67*,4961-4980.   http://doi.org/10.1016/s0016-7037(03)00485-x.

38. Timmes, F.X.; Clayton, D.D. Galactic evolution of silicon isotopes: application to presolar SiC grains from meteorites. *Astrophys. J.* **1996**, *472*,723.   http://doi.org/10.1086/178102.

39. Lewis, K.M.; Lugaro, M.; Gibson, B.K.; Pilkington, K. Decoding the message from meteoritic stardust silicon carbide grains. *Astrophys. J.* **2013**, *768*,L19.   http://doi.org/10.1088/2041-8205/768/1/l19.

40. Cristallo, S.; Nanni, A.; Cescutti, G.; Minchev, I.; Liu, N.; Vescovi, D.; Gobrecht, D.; Piersanti, L. Mass and metallicity distribution of parent AGB stars of presolar SiC. *Astron. Astrophys.* **2020**, *644*, A8. http://doi.org/10.1051/0004-6361/202039492.
41. Liu, N.; Gallino, R.; Cristallo, S.; Bisterzo, S.; Davis, A.M.; Trappitsch, R.; Nittler, L.R. New constraints on the major neutron source in low-mass AGB stars. *Astrophys. J.* **2018**, *865*, 112. http://doi.org/10.3847/1538-4357/aad9f3.
42. Lugaro, M.; Karakas, A.I.; Pető, M.; Plachy, E. Do meteoritic silicon carbide grains originate from asymptotic giant branch stars of super-solar metallicity? *Geochim. Cosmochim. Acta* **2018**, *221*, 6-20. http://doi.org/10.1016/j.gca.2017.06.006.
43. Lugaro, M.; Cseh, B.; Világos, B.; Karakas, A.I.; Ventura, P.; Dell'agli, F.; Trappitsch, R.; Hampel, M.; D'orazi, V.; Pereira, C.B.; Tagliente, G.; Szabó, G.M.; Pignatari, M.; Battino, U.; Tattersall, A.; Ek, M.; Schönbächler, M.; Hron, J.; Nittler, L.R. Origin of large meteoritic SiC stardust grains in metal-rich AGB stars. *Astrophys. J.* **2020**, *898*, 96. http://doi.org/10.3847/1538-4357/ab9e74.
44. Amari, S.; Nittler, L.R.; Zinner, E.; Gallino, R.; Lugaro, M.; Lewis, R.S. Presolar SiC grains of type Y: Origin from low-metallicity asymptotic giant branch stars. *Astrophys. J.* **2001**, *546*, 248-266. http://doi.org/10.1086/318230.
45. Zinner, E.; Nittler, L.R.; Gallino, R.; Karakas, A.I.; Lugaro, M.; Straniero, O.; Lattanzio, J.C. Silicon and carbon isotopic ratios in AGB stars: SiC grain data, models, and the Galactic evolution of the Si isotopes. *Astrophys. J.* **2006**, *650*, 350-373. http://doi.org/10.1086/506957.
46. Liu, N.; Stephan, T.; Cristallo, S.; Gallino, R.; Boehnke, P.; Nittler, L.R.; O'd. Alexander, C.M.; Davis, A.M.; Trappitsch, R.; Pellin, M.J. Presolar silicon carbide grains of groups Y and Z: their strontium and barium isotopic compositions and stellar origins. 50th Lunar and Planetary Science Conference, Houston, TX, March 01, 2019, 1349 **2019**.
47. Liu, N.; Stephan, T.; Cristallo, S.; Gallino, R.; Boehnke, P.; Nittler, L.R.; O'd. Alexander, C.M.; Davis, A.M.; Trappitsch, R.; Pellin, M.J.; Dillmann, I. Presolar silicon carbide grains of types Y and Z: their molybdenum isotopic compositions and stellar origins. *Astrophys. J.* **2019**, *881*, 28. http://doi.org/10.3847/1538-4357/ab2d27.
48. Boujibar, A.; Howell, S.; Zhang, S.; Hystad, G.; Prabhu, A.; Liu, N.; Stephan, T.; Narkar, S.; Eleish, A.; Morrison, S.M.; Hazen, R.M.; Nittler, L.R. Cluster analysis of presolar silicon carbide grains: evaluation of their classification and astrophysical implications. *Astrophys. J.* **2021**, *907*, L39. http://doi.org/10.3847/2041-8213/abd102.
49. Hystad, G.; Boujibar, A.; Liu, N.; Nittler, L.R.; Hazen, R.M. Evaluation of the classification of pre-solar silicon carbide grains using consensus clustering with resampling methods: An assessment of the confidence of grain assignments. *Mon. Not. R. Astron. Soc.* **2022**, *510*, 334-350. http://doi.org/10.1093/mnras/stab3478.
50. Liu, N.; Barosch, J.; Nittler, L.R.; O'd. Alexander, C.M.; Wang, J.; Cristallo, S.; Busso, M.; Palmerini, S. New multielement isotopic compositions of presolar SiC grains: implications for their stellar origins. *Astrophys. J.* **2021**, *920*, L26. http://doi.org/10.3847/2041-8213/ac260b.
51. Liu, N.; M., O.D.a.C.; Nittler, L.R. Intrinsic nitrogen isotope ratios of presolar silicon carbide grains. 85th Annual Meeting of The Meteoritical Society, Glascow, U.K. 6384 **2022**.
52. Stephan, T.; Bose, M.; Boujibar, A.; Davis, A.M.; Gyngard, F.; Hoppe, P.; Hynes, K.M.; Liu, N.; Nittler, L.R.; Ogliore, R.C.; Trappitsch, R. The Presolar Grain Database for silicon carbide — grain type assignments. 52nd Lunar and Planetary Science Conference, Houston, TX, March 01, 2021, 2358 **2021**.
53. Hoppe, P.; Leitner, J.; Gröner, E.; Marhas, K.K.; Meyer, B.S.; Amari, S. NanoSIMS studies of small presolar SiC grains: new insights into supernova nucleosynthesis, chemistry, and dust formation. *Astrophys. J.* **2010**, *719*, 1370-1384. http://doi.org/10.1088/0004-637x/719/2/1370.
54. Leitner, J.; Hoppe, P. A new population of dust from stellar explosions among meteoritic stardust. *Nat. Astron.* **2019**, *3*, 725-729. http://doi.org/10.1038/s41550-019-0788-x.

55. Nittler, L.R.; O'd. Alexander, C.M.; Liu, N.; Wang, J. Extremely $^{54}$Cr- and $^{50}$Ti-rich presolar oxide grains in a primitive meteorite: Formation in rare types of supernovae and implications for the astrophysical context of solar system birth. *Astrophys. J.* **2018**, *856*,L24.   http://doi.org/10.3847/2041-8213/aab61f.

56. Liu, N.; Dauphas, N.; Cristallo, S.; Palmerini, S.; Busso, M. Oxygen and aluminum-magnesium isotopic systematics of presolar nanospinel grains from CI chondrite Orgueil. *Geochim. Cosmochim. Acta* **2022**, *319*,296-317. http://doi.org/10.1016/j.gca.2021.11.022.

57. Lodders, K. Relative atomic solar system abundances, mass fractions, and atomic masses of the elements and their isotopes, composition of the solar photosphere, and compositions of the major chondritic meteorite groups. *Space Sci. Rev.* **2021**, *217*,44.   http://doi.org/10.1007/s11214-021-00825-8.

58. Amari, S.; Hoppe, P.; Zinner, E.; Lewis, R.S. Trace-element concentrations in single circumstellar silicon carbide grains from the Murchison meteorite. *Meteoritics* **1995**, *30*,679.   http://doi.org/10.1111/j.1945-5100.1995.tb01165.x.

59. Trappitsch, R.; Ong, W.-J.; Dory, C.J.; Shulaker, D.Z.; Lugaro, M.; Savina, M.R.; Weber, P.K.; Isselhardt, B.H.; Amari, S. Simultaneous analyses of titanium and molybdenum isotopic compositions in presolar SiC grains. 84th Annual Meeting of the Meteoritical Society, August 01, 2021, 6239 **2021**.

60. Trappitsch, R.; Stephan, T.; Savina, M.R.; Davis, A.M.; Pellin, M.J.; Rost, D.; Gyngard, F.; Gallino, R.; Bisterzo, S.; Cristallo, S.; Dauphas, N. Simultaneous iron and nickel isotopic analyses of presolar silicon carbide grains. *Geochim. Cosmochim. Acta* **2018**, *221*,87-108.   http://doi.org/10.1016/j.gca.2017.05.031.

61. Liu, N.; Savina, M.R.; Gallino, R.; Davis, A.M.; Bisterzo, S.; Gyngard, F.; Käppeler, F.; Cristallo, S.; Dauphas, N.; Pellin, M.J.; Dillmann, I. Correlated strontium and barium isotopic compositions of acid-cleaned single mainstream silicon carbides from Murchison. *Astrophys. J.* **2015**, *803*,12.   http://doi.org/10.1088/0004-637x/803/1/12.

62. Liu, N.; Savina, R.M.; Davis, A.M.; Willingham, D.G.; Pellin, M.; Dauphas, N. Barium and neodymium isotopic composition of presolar SiC grains. *Meteoritics and Planetary Science Supplement* **2012**, *75*,5249.

63. Barzyk, J.G.; Savina, M.R.; Davis, A.M.; Gallino, R.; Gyngard, F.; Amari, S.; Zinner, E.; Pellin, M.J.; Lewis, R.S.; Clayton, R.N. Constraining the $^{13}$C neutron source in AGB stars through isotopic analysis of trace elements in presolar SiC. *Meteorit. Planet. Sci.* **2007**, *42*,1103-1119.   http://doi.org/10.1111/j.1945-5100.2007.tb00563.x.

64. Nicolussi, G.K.; Pellin, M.J.; Lewis, R.S.; Davis, A.M.; Clayton, R.N.; Amari, S. Strontium isotopic composition in individual circumstellar silicon carbide grains: a record of s-process nucleosynthesis. *Phys. Rev. Lett.* **1998**, *81*,3583-3586. http://doi.org/10.1103/PhysRevLett.81.3583.

65. Nicolussi, G.K.; Pellin, M.J.; Lewis, R.S.; Davis, A.M.; Amari, S.; Clayton, R.N. Molybdenum isotopic composition of individual presolar silicon carbide grains from the Murchison meteorite. *Geochim. Cosmochim. Acta* **1998**, *62*,1093-1104. http://doi.org/10.1016/s0016-7037(98)00038-6.

66. Stephan, T.; Trappitsch, R.; Hoppe, P.; Davis, A.M.; Pellin, M.J.; Pardo, O.S. Molybdenum isotopes in presolar silicon carbide grains: details of s-process nucleosynthesis in parent stars and implications for r- and p-processes. *Astrophys. J.* **2019**, *877*,101. http://doi.org/10.3847/1538-4357/ab1c60.

67. Savina, M.R.; Davis, A.M.; Tripa, C.E.; Pellin, M.J.; Gallino, R.; Lewis, R.S.; Amari, S. Extinct technetium in silicon carbide stardust grains: implications for stellar nucleosynthesis. *Science* **2004**, *303*,649-652.   http://doi.org/10.1126/science.3030649.

68. Stephan, T.; Trappitsch, R.; Davis, A.M.; Pellin, M.J.; Rost, D.; Savina, M.R.; Yokochi, R.; Liu, N. CHILI - the Chicago Instrument for Laser Ionization - a new tool for isotope measurements in cosmochemistry. *Int. J. Mass Spectrom.* **2016**, *407*,1-15.   http://doi.org/10.1016/j.ijms.2016.06.001.

69. Savina, M.R.; Pellin, M.J.; Tripa, C.E.; Veryovkin, I.V.; Calaway, W.F.; Davis, A.M. Analyzing individual presolar grains with CHARISMA. *Geochim. Cosmochim. Acta* **2003**, *67*,3215-3225.   http://doi.org/10.1016/s0016-7037(03)00082-6.

70. Clayton, D.D.; Fowler, W.A.; Hull, T.E.; Zimmerman, B.A. Neutron capture chains in heavy element synthesis. *Annals of Physics* **1961**, *12*,331-408. http://doi.org/10.1016/0003-4916(61)90067-7.
71. Arlandini, C.; Käppeler, F.; Wisshak, K.; Gallino, R.; Lugaro, M.; Busso, M.; Straniero, O. Neutron capture in low-mass asymptotic giant branch stars: cross sections and abundance signatures. *Astrophys. J.* **1999**, *525*,886-900. http://doi.org/10.1086/307938.
72. Clayton, D.D.; Ward, R.A. S-process studies: exact evaluation of an exponential distribution of exposures. *Astrophys. J.* **1974**, *193*,397-400. http://doi.org/10.1086/153175.
73. Busso, M.; Gallino, R.; Lambert, D.L.; Travaglio, C.; Smith, V.V. Nucleosynthesis and mixing on the asymptotic giant branch. III. predicted and observed *s*-process abundances. *Astrophys. J.* **2001**, *557*,802-821. http://doi.org/10.1086/322258.
74. Cristallo, S.; Piersanti, L.; Straniero, O.; Gallino, R.; Domínguez, I.; Abia, C.; Di Rico, G.; Quintini, M.; Bisterzo, S. Evolution, nucleosynthesis, and yields of low-mass asymptotic giant branch stars at different metallicities. II. The FRUITY database. *Astrophys. J. Suppl. Ser.* **2011**, *197*,17. http://doi.org/10.1088/0067-0049/197/2/17.
75. Vescovi, D.; Cristallo, S.; Busso, M.; Liu, N. Magnetic-buoyancy-induced mixing in AGB stars: presolar SiC grains. *Astrophys. J.* **2020**, *897*,L25. http://doi.org/10.3847/2041-8213/ab9fa1.
76. Sneden, C.; Cowan, J.J.; Gallino, R. Neutron-capture elements in the early Galaxy. *Annu. Rev. Astron. Astrophys.* **2008**, *46*,241-288. http://doi.org/10.1146/annurev.astro.46.060407.145207.
77. Lambert, D.L.; Allende Prieto, C. The isotopic mixture of barium in the metal-poor subgiant HD 140283. *Mon. Not. R. Astron. Soc.* **2002**, *335*,325-334. http://doi.org/10.1046/j.1365-8711.2002.05643.x.
78. Stephan, T.; Trappitsch, R.; Davis, A.M.; Pellin, M.J.; Rost, D.; Savina, M.R.; Jadhav, M.; Kelly, C.H.; Gyngard, F.; Hoppe, P.; Dauphas, N. Strontium and barium isotopes in presolar silicon carbide grains measured with CHILI-two types of X grains. *Geochim. Cosmochim. Acta* **2018**, *221*,109-126. http://doi.org/10.1016/j.gca.2017.05.001.
79. Savina, M.R.; Davis, A.M.; Tripa, C.E.; Pellin, M.J.; Clayton, R.N.; Lewis, R.S.; Amari, S.; Gallino, R.; Lugaro, M. Barium isotopes in individual presolar silicon carbide grains from the Murchison meteorite. *Geochim. Cosmochim. Acta* **2003**, *67*,3201-3214. http://doi.org/10.1016/s0016-7037(03)00083-8.
80. Takahashi, K.; Yokoi, K. Beta-decay rates of highly ionized heavy atoms in stellar interiors. *At. Data Nucl. Data Tables* **1987**, *36*,375. http://doi.org/10.1016/0092-640x(87)90010-6.
81. Longland, R.; Iliadis, C.; Karakas, A.I. Reaction rates for the s-process neutron source $^{22}$Ne + $\alpha$. *Phys. Rev. C* **2012**, *85*,065809. http://doi.org/10.1103/PhysRevC.85.065809.
82. Busso, M.; Gallino, R.; Wasserburg, G.J. Nucleosynthesis in asymptotic giant branch stars: relevance for Galactic enrichment and solar system formation. *Annu. Rev. Astron. Astrophys.* **1999**, *37*,239-309. http://doi.org/10.1146/annurev.astro.37.1.239.
83. Vescovi, D.; Cristallo, S.; Palmerini, S.; Abia, C.; Busso, M. Magnetic-buoyancy-induced mixing in AGB stars: fluorine nucleosynthesis at different metallicities. *Astron. Astrophys.* **2021**, *652*,A100. http://doi.org/10.1051/0004-6361/202141173.
84. Vescovi, D.; Piersanti, L.; Cristallo, S.; Busso, M.; Vissani, F.; Palmerini, S.; Simonucci, S.; Taioli, S. Effects of a revised $^7$Be e-capture rate on solar neutrino fluxes. *Astron. Astrophys.* **2019**, *623*,A126. http://doi.org/10.1051/0004-6361/201834993.
85. Vescovi, D.; Mascaretti, C.; Vissani, F.; Piersanti, L.; Straniero, O. The luminosity constraint in the era of precision solar physics. *J. Phys. G: Nucl. Part. Phys.* **2021**, *48*,015201. http://doi.org/10.1088/1361-6471/abb784.
86. Adsley, P.; Battino, U.; Best, A.; Caciolli, A.; Guglielmetti, A.; Imbriani, G.; Jayatissa, H.; La Cognata, M.; Lamia, L.; Masha, E.; Massimi, C.; Palmerini, S.; Tattersall, A.; Hirschi, R. Reevaluation of the $^{22}$Ne($\alpha,\gamma$)$^{26}$Mg and $^{22}$Ne($\alpha,n$)$^{25}$Mg reaction rates. *Phys. Rev. C* **2021**, *103*,015805. http://doi.org/10.1103/PhysRevC.103.015805.
87. Bao, Z.Y.; Beer, H.; Käppeler, F.; Voss, F.; Wisshak, K.; Rauscher, T. Neutron cross sections for nucleosynthesis studies. *At. Data Nucl. Data Tables* **2000**, *76*,70-154. http://doi.org/10.1006/adnd.2000.0838.

88. Taioli, S.; Vescovi, D.; Busso, M.; Palmerini, S.; Cristallo, S.; Mengoni, A.; Simonucci, S. 2021. Theoretical estimate of the half-life for the radioactive $^{134}$Cs and $^{135}$Cs in astrophysical scenarios. arXiv:2109.14230. Accessed September 01, 2021.
89. Patronis, N.; Dababneh, S.; Assimakopoulos, P.A.; Gallino, R.; Heil, M.; Käppeler, F.; Karamanis, D.; Koehler, P.E.; Mengoni, A.; Plag, R. Neutron capture studies on unstable $^{135}$Cs for nucleosynthesis and transmutation. *Phys. Rev. C* **2004**, *69*,025803. http://doi.org/10.1103/PhysRevC.69.025803.
90. Reifarth, R.; Erbacher, P.; Fiebiger, S.; Göbel, K.; Heftrich, T.; Heil, M.; Käppeler, F.; Klapper, N.; Kurtulgil, D.; Langer, C.; Lederer-Woods, C.; Mengoni, A.; Thomas, B.; Schmidt, S.; Weigand, M.; Wiescher, M. Neutron-induced cross sections. From raw data to astrophysical rates. *Eur. Phys. J. Plus* **2018**, *133*,424. http://doi.org/10.1140/epjp/i2018-12295-3.
91. Jayatissa, H.; Rogachev, G.V.; Goldberg, V.Z.; Koshchiy, E.; Christian, G.; Hooker, J.; Ota, S.; Roeder, B.T.; Saastamoinen, A.; Trippella, O.; Upadhyayula, S.; Uberseder, E. Constraining the $^{22}$Ne($\alpha,\gamma$)$^{26}$Mg and $^{22}$Ne($\alpha$,n)$^{25}$Mg reaction rates using sub-Coulomb $\alpha$-transfer reactions. *Phys. Lett. B* **2020**, *802*,135267. http://doi.org/10.1016/j.physletb.2020.135267.
92. Li, K.-A.; Qi, C.; Lugaro, M.; Yagüe López, A.; Karakas, A.I.; Den Hartogh, J.; Gao, B.-S.; Tang, X.-D. The stellar β-decay rate of $^{134}$Cs and its impact on the barium nucleosynthesis in the s-process. *Astrophys. J.* **2021**, *919*,L19. http://doi.org/10.3847/2041-8213/ac260f.
93. Palmerini, S.; Busso, M.; Vescovi, D.; Naselli, E.; Pidatella, A.; Mucciola, R.; Cristallo, S.; Mascali, D.; Mengoni, A.; Simonucci, S.; Taioli, S. Presolar grain isotopic ratios as constraints to nuclear and stellar parameters of asymptotic giant branch star nucleosynthesis. *Astrophys. J.* **2021**, *921*,7. http://doi.org/10.3847/1538-4357/ac1786.
94. Busso, M.; Vescovi, D.; Palmerini, S.; Cristallo, S.; Antonuccio-Delogu, V. s-processing in AGB stars revisited. III. Neutron captures from MHD mixing at different metallicities and observational constraints. *Astrophys. J.* **2021**, *908*,55. http://doi.org/10.3847/1538-4357/abca8e.
95. Dillmann, I. The new KADoNiS v1.0 and its influence on the s-process. XIII Nuclei in the Cosmos (NIC XIII), January 01, 2014, 57 **2014**.
96. Sasaki, H.; Yamazaki, Y.; Kajino, T.; Kusakabe, M.; Hayakawa, T.; Cheoun, M.-K.; Ko, H.; Mathews, G.J. Impact of hypernova νp-process nucleosynthesis on the Galactic chemical evolution of Mo and Ru. *Astrophys. J.* **2022**, *924*,29. http://doi.org/10.3847/1538-4357/ac34f8.
97. Lugaro, M.; Davis, A.M.; Gallino, R.; Pellin, M.J.; Straniero, O.; Käppeler, F. Isotopic compositions of strontium, zirconium, molybdenum, and barium in single presolar SiC grains and asymptotic giant branch stars. *Astrophys. J.* **2003**, *593*,486-508. http://doi.org/10.1086/376442.
98. Pignatari, M.; Gallino, R.; Heil, M.; Wiescher, M.; Käppeler, F.; Herwig, F.; Bisterzo, S. The weak s-process in massive stars and its dependence on the neutron capture cross sections. *Astrophys. J.* **2010**, *710*,1557-1577. http://doi.org/10.1088/0004-637x/710/2/1557.